\documentclass[11pt]{article}

\usepackage[a4paper, margin=2.5cm]{geometry}
\usepackage{amsmath, amssymb, amsthm}
\usepackage{graphicx}
\usepackage[english]{babel}
\usepackage[utf8]{inputenc}
\usepackage{xcolor}
\usepackage{tabularx}
\usepackage{siunitx}
\usepackage{multirow}

\usepackage{hyperref}
\hypersetup{
	colorlinks=true,
	linkcolor=blue,
	citecolor=blue,
	urlcolor=blue
}

\begin{document}
	
\interfootnotelinepenalty=10000
	
\title{ADOPT: Analytical Demodulation of Periodic Textures for In-Plane Wave Tracking}
\author{J. Jouidi, F. Chatelain, L. Bailly, P. Granjon, N. Le Bihan, S. Catheline}
\maketitle

\begin{abstract}
    \noindent
This paper addresses the problem of tracking in-plane waves from image sequences using periodic surface patterns. Wave-induced deformation is modeled as a spatial phase modulation of a periodic carrier. We propose ADOPT (Analytical Demodulation of Periodic Texture), a method based on an oriented two-dimensional analytic signal to estimate displacement phase and orientation.

The approach relies on a physical model describing longitudinal and transverse in-plane waves. Orientation-selective filtering isolates relevant spectral components, and phase extraction provides a stable reconstruction of the displacement field.

A theoretical analysis using the Cramér--Rao bound evaluates performance limits of ADOPT. Simulations show that the proposed method outperforms state-of-the-art Digital Image Correlation (DIC) at high signal-to-noise ratios, especially for small displacements where DIC becomes limited. Moreover, ADOPT is more computationally efficient.

Experiments on silicone membranes with periodic patterns confirm accurate estimation of wave fields and dispersion curves under impulsive excitation. Overall, the proposed framework provides a robust and efficient solution for wave-induced displacement estimation.
\end{abstract}

\section{\label{sec:1} Introduction}

Non-contact imaging methods are widely employed to characterize displacement and strain fields under dynamic excitation, with applications ranging from vibration analysis to structural health monitoring. Among them, Digital Image Correlation (DIC) has served as a benchmark technique since the 1980s \cite{rastogi2003photomechanics}\cite[Sec.~6.2]{grediac2012full}. More generally, DIC belongs to the class of optical flow approaches, which estimate motion by correlating local image subsets. Advances in high-speed cameras now allow transient phenomena to be resolved at microsecond time scales \cite{besnard2010characterization}. These techniques span multiple spatial scales, from nanometric measurements using atomic force microscopy \cite{chasiotis2002new}, to macroscopic optical imaging \cite{lanoy2020dirac}, satellite observations \cite{leprince2007automatic}, biological \cite{grasland2018ultrafast}, and medical applications \cite{sutton2008strain}. The resulting full-field kinematic measurements are commonly used for model identification and estimation of elastic or viscoelastic parameters \cite{zhu2025young}.

While deep learning strategies have been proposed for optical flow estimation, their dependence on large training datasets and limited robustness hinder deployment in experimental settings. As a result, model-based signal processing methods remain preferable for practical applications \cite{alfarano2024estimating}.

\begin{figure}[h!]
    \centering
    \includegraphics[height=5cm]{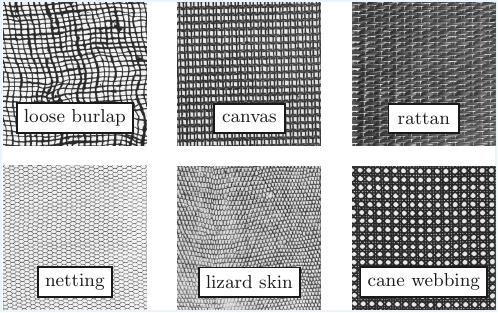}
    \caption{Examples of periodic surface texture patterns extracted from the standard Brodatz grayscale texture album \cite{brodatz1966textures}.}
    \label{fig:outex}
\end{figure}

In this work, surface waves are tracked from image sequences using periodic surface patterns. In contrast to classical DIC, which relies on random speckle textures, periodic patterns either naturally exist in materials (Fig.~\ref{fig:outex}) or can be printed on specimens. Wave-induced motion can then be interpreted as a phase modulation of the underlying pattern. Exploiting periodicity alleviates the limitations of DIC on repetitive textures, notably those arising from the aperture problem and the absence of distinctive features. Beyond this specific configuration, the proposed method consistently improves performance across all tested scenarios, making periodic patterns combined with our framework advantageous whenever feasible.

Displacements are estimated relative to a reference image, without explicit temporal modeling. The method couples the periodic pattern with a demodulation scheme based on a modified two-dimensional analytic signal, enabling joint recovery of displacement phase and orientation. Both longitudinal and transverse in-plane waves are handled, extending previous work \cite{wildeman2018real,domino2016faraday} through a generalized modulation model, tighter theoretical bounds, and explicit orientation estimation.

The main contributions are twofold: (i) a physics-informed parametric signal model tailored to periodic structures, together with an inverse formulation for retrieving small-amplitude wave fields from deformed patterns using an appropriate analytic signal; and (ii) a combined theoretical performance analysis via the Cramér–Rao bound and a quantitative comparison with state-of-the-art DIC under controlled conditions. To the authors’ knowledge, this unified modeling, analysis, and benchmarking framework has not been reported previously.

Section~\ref{sec:mathmodel} introduces the modulation model, and Section~\ref{sec:demod} details the demodulation strategy. Section~\ref{sec:results} presents numerical simulations, where noise levels and pattern properties are controlled to assess performance in comparison with DIC and robustness to noise. Experimental validation is then conducted in Section~\ref{sec:exp} on thin silicone membranes with printed surface patterns, where dispersion curves of longitudinal in-plane waves are estimated using both approaches. The results highlight the improved robustness of the proposed method, particularly for small displacements.

\section{Mathematical description of wave-induced modulation of periodic surface patterns}
\label{sec:mathmodel}

\begin{figure}[h!]
    \centering
    \includegraphics[height=5.5cm]{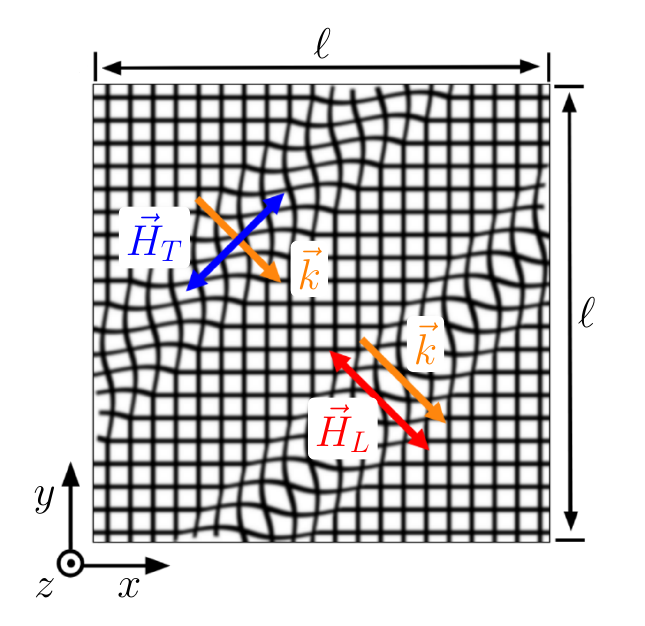}
    \caption{In-plane wave propagation (zero out-of-plane displacement) in a square membrane of side length $\ell$. Longitudinal $\vec{H}_L$ and transverse $\vec{H}_T$ components are shown. The vector $\vec{k}$ denotes the propagation direction.} 
    \label{fig:waves}
\end{figure}

In this study, only in-plane displacements are considered, while out-of-plane motion is neglected. We consider a two-dimensional displacement field $\vec{H}=(U,V)$ evolving within the plane of a thin membrane, where $U,V:\mathbb{R}^2 \to \mathbb{R}$ are scalar fields denoting the displacement components along $x$ and $y$, respectively. 

According to the Helmholtz decomposition theorem \cite[Sec.~5.1]{graff2012wave}, such a vector field can be uniquely decomposed into the sum of an irrotational (longitudinal) part and a divergence-free (transverse) part:
\begin{equation}
\vec{H} = \vec{H}_L + \vec{H}_T,
\end{equation}
where $\vec{H}_L = (U_L, V_L)$ and $\vec{H}_T = (U_T, V_T)$. 
In this framework, longitudinal waves $\vec{H}_L$ induce particle motion parallel to the propagation direction $\vec{k}$, whereas transverse waves $\vec{H}_T$ generate motion orthogonal to $\vec{k}$, as illustrated in Fig.~\ref{fig:waves}.
The longitudinal and transverse components can be expressed in terms of potentials $P_L$ and $P_T$, respectively, such that

\begin{equation}
U_L = \frac{\partial P_L}{\partial x}, \quad V_L = \frac{\partial P_L}{\partial y},
\end{equation}

\begin{equation}
U_T = \frac{\partial P_T}{\partial y}, \quad V_T = -\frac{\partial P_T}{\partial x}.
\end{equation}

The total displacement field is then obtained by summing the two components:
\begin{equation}
U = \frac{\partial P_L}{\partial x} + \frac{\partial P_T}{\partial y}, \quad
V =  \frac{\partial P_L}{\partial y} - \frac{\partial P_T}{\partial x},
\label{eq:UV}
\end{equation}
and where $P_L, P_T: \mathbb{R}^2 \to \mathbb{R}$ are scalar fields. 
These fields are assumed to be smooth with respect to $x$ and $y$, bounded in amplitude by a constant $a_p$, i.e., $|P_L(x,y)|, |P_T(x,y)| \leq a_p$, and band-limited in spatial frequency by the vector $\vec{\xi}_m = (\xi_{mx}, \xi_{my})$, which magnitude, denoted $\xi_m$, is given by
\[
\xi_m = \sqrt{\xi_{mx}^2 + \xi_{my}^2}.
\]
The use of these potentials allows the simulation of longitudinal and transverse waves. Partial derivatives yield the corresponding components and ensure the curl-free and divergence-free conditions.

Let $I(x,y)$ denote the reference surface pattern. Under wave propagation, the observed surface is deformed. Introducing the mapping $\Phi:\mathbb{R}^2 \to \mathbb{R}^2$ defined as
\begin{equation}
\Phi(x,y) = \big(x - U(x,y),\, y - V(x,y)\big),
\end{equation}
where $U$ and $V$ are given in Equation~\eqref{eq:UV}.
The deformed surface pattern can be written as
\begin{equation}
I_\textrm{d}(x,y)=I\big(\Phi(x,y)\big).
\label{eq:Fmod}
\end{equation}
The mapping $\Phi$ therefore characterizes the deformation of the membrane. For clarity, the explicit dependence on $(x,y)$ is omitted in the remainder of this section. The following analysis focuses on some of the physical constraints that will be imposed in the proposed model.

\subsection{Physical constraints}

From the constraints previously imposed on the potentials, one can infer, using Bernstein's inequality \footnote{Bernstein's inequality states that if $f: \mathbb{R}^2 \to \mathbb{R}$ is a scalar field such that $|f(x,y)| \leq a$ and its Fourier transform is band-limited by $\vec{\xi}_m = (\xi_{mx}, \xi_{my})$ whose magnitude is $\xi_m = \sqrt{\xi_{mx}^2 + \xi_{my}^2}$, then its spatial derivatives are also bounded. In particular, it holds that $\left|\frac{\partial f}{\partial x}\right| \leq 2\pi a \xi_m$ and $\left|\frac{\partial f}{\partial y}\right| \leq 2\pi a \xi_m$.} \cite[Sec.~2.3.8]{pinsky2008introduction}, the corresponding constraints on the displacement fields $U$ and $V$. 
Specifically, the displacements are bounded in amplitude by $a$, with the amplitude related to the potential bound $a_p$ through
\[
a = 2\pi \xi_m a_p,
\]
and band-limited in spatial frequency by $\vec{\xi}_m$, consistent with the potentials.

A fundamental physical constraint relating the magnitude of the maximum spatial frequency $\xi_m$ to the displacement amplitude $a$ is that the mapping $\Phi$ must be bijective and maintain the orientation of the surface at every point.  
Bijectivity guarantees a one-to-one mapping between points of the original and deformed patterns, while preserving orientation prevents sudden folding or inversion, which a physical membrane cannot undergo without rupturing.

Mathematically, this constraint can be expressed as a positivity condition on the determinant of the Jacobian matrix of $\Phi$ at every point $(x,y)$ on the surface:
\begin{equation}
\det 
\begin{pmatrix}
1 - \frac{\partial U}{\partial x} & -\frac{\partial U}{\partial y} \\
-\frac{\partial V}{\partial x} & 1 - \frac{\partial V}{\partial y}
\end{pmatrix} > 0.
\end{equation}

Since $U$ and $V$ are band-limited by $\xi_m$, Bernstein's inequality gives an upper bound on their derivatives:
\begin{equation}
\left| \frac{\partial U}{\partial x} \right|, 
\left| \frac{\partial U}{\partial y} \right|,
\left| \frac{\partial V}{\partial x} \right|,
\left| \frac{\partial V}{\partial y} \right| \le 2\pi a \, \xi_m.
\label{eq:born_partial}
\end{equation}

Thus, to guarantee that $\Phi$ remains bijective and orientation-preserving, the displacement amplitude and spatial bandwidth must satisfy
\begin{equation}
\xi_m < \frac{1}{4\pi a}.
\label{eq:physical_const}
\end{equation}

This inequality represents a key constraint for our model. For a given displacement amplitude $a$, it sets an upper limit on the maximum spatial frequency $\xi_m$, thereby reducing the degrees of freedom of the model and allowing it to be fully characterized using only the maximum displacement $a$. 

In the subsequent analysis, we investigate the spectral consequences of deformations induced by waves on the surface, focusing on the specific case of a periodic pattern.

\subsection{Fourier representation of a periodic pattern}
\label{sec:pattern}

\begin{figure}[h!]
     \centering
     \includegraphics[width= 9cm,trim=30 90 0 0, clip]{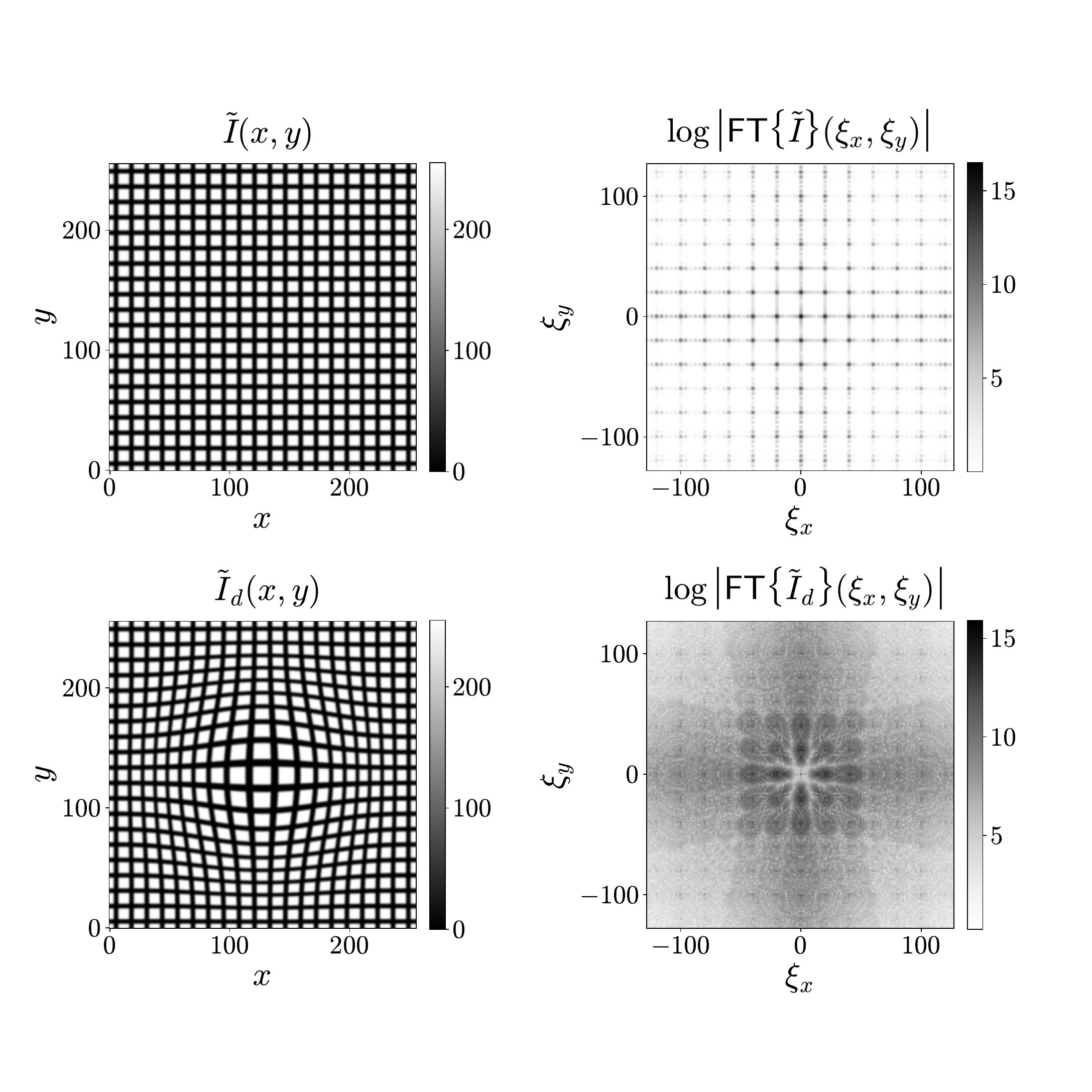}
     \caption{Pattern on the surface and corresponding 2D Fourier transform. \textbf{Left:} pattern $\tilde{I}$ at rest (top) and submitted to a deformation $\tilde{I}_\textrm{d}$ (bottom). \textbf{Right:} 2D Fourier transform (magnitude in log scale) of $\tilde{I}$ (top) and of $\tilde{I}_\textrm{d}$ (bottom). The colormap is inverted in the frequency domain for better visualization.}
     \label{fig:pattern_FT}
\end{figure}

We assume that the membrane is textured with a pattern that is periodic along both spatial directions, with fundamental frequency $\xi_p$. Let $e(x,y)$ denote a compact-support elementary image of side length $\epsilon$, which is periodically replicated to generate the full pattern. For instance, in Fig.~\ref{fig:pattern_FT}, $e$ corresponds to \includegraphics[height=2ex]{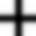}. This replication process can be modeled using a two-dimensional Dirac comb,
\begin{equation}
\Lambda(x,y)=\sum_{\gamma,\kappa\in\mathbb{Z}}\delta(x-\gamma\epsilon,\,y-\kappa\epsilon),
\end{equation}
where $\epsilon$ represents the lattice spacing and satisfies $\epsilon=1/\xi_p$. The resulting pattern is expressed as $I(x,y)=e(x,y)*\Lambda(x,y)$, with $*$ denoting two-dimensional convolution.

Taking the Fourier transform yields
\begin{equation}
\mathsf{FT}\{I\}(\xi_x,\xi_y)=
\mathsf{FT}\{e\}(\xi_x,\xi_y)\frac{1}{\xi_p^2}
\sum_{\gamma,\kappa\in\mathbb{Z}}
\delta(\xi_x-\gamma\xi_p,\,\xi_y-\kappa\xi_p),
\label{eq:FT_no_mod}
\end{equation}
revealing a lattice of spectral impulses weighted by the broad spectrum of $e(x,y)$, since the unit cell is much smaller than the membrane size ($\epsilon\ll\ell$). Consequently, $\mathsf{FT}\{e\}(\xi_x,\xi_y)$ mainly shapes the amplitudes of the harmonic peaks (Fig.~\ref{fig:pattern_FT}, top right).

Introducing the displacement field through \eqref{eq:Fmod} produces spectral spreading around each harmonic. The lattice spacing $\epsilon$ plays a key role in controlling this effect. A square unit cell minimizes spectral overlap. Combining \eqref{eq:Fmod} and \eqref{eq:FT_no_mod}, the Fourier transform of the modulated pattern becomes
\begin{align}
\mathsf{FT}\{I_\textrm{d}\}(\xi_x,\xi_y)
&=\mathsf{FT}\{e\}(\xi_x,\xi_y)\nonumber\\
&\quad \frac{1}{\xi_p^2}\sum_{\gamma,\kappa\in\mathbb{Z}}
\delta(\xi_x-\gamma\xi_p,\,\xi_y-\kappa\xi_p)\nonumber\\
&\quad *\,\mathsf{FT}\!\left\{e^{-2\pi i\xi_p(\gamma U+\kappa V)}\right\}(\xi_x,\xi_y).
\label{eq:FT_Fmod}
\end{align}

Each spectral harmonic is therefore convolved with the spectrum of $e^{-2\pi i\xi_p(\gamma U+\kappa V)}$, which theoretically has infinite support (Fig.~\ref{fig:pattern_FT}, bottom right).

In practice, most of the energy remains confined within a finite bandwidth. Empirical criteria, mainly used in the telecommunications domain, such as Carson’s rule \cite{carson2006notes}, provides an effective estimate capturing approximately 98\% of the signal power. In the one-dimensional case, Carson’s rule states that the bandwidth $c$ of a frequency modulated signal is
\[
c = 2(\Delta \xi + \xi_m),
\]
where $\Delta \xi$ is the peak frequency deviation and $\xi_m$ is the highest frequency in the modulating signal.

Note that while modulation in telecommunications is typically with respect to time, in our case it occurs in space, resulting in a spatial frequency modulated signal.

To extend the notion of Carson's rule to 2D signals, we need to generalize the concept of peak frequency deviation to the two-dimensional modulated case. For a given harmonic $(\gamma,\kappa) \in \mathbb{Z}^2$, the peak frequency deviation along $x$ and $y$ can be written as $\xi_p \gamma \max\left(\left| \frac{\partial U}{\partial x} \right|\right)$ and $\xi_p \kappa \max\left(\left| \frac{\partial V}{\partial y} \right|\right)$, respectively. 

Using the bound established in \eqref{eq:born_partial}, the deviations along $x$ and $y$ become $2\pi a \xi_p \xi_m \gamma$ and $2\pi a \xi_p \xi_m \kappa$, respectively. 

This makes it possible to define a peak frequency deviation vector: $(2\pi a \xi_p \xi_m \gamma,\; 2\pi a \xi_p \xi_m \kappa)$, whose norm gives the resulting peak deviation in 2D. Thus, Carson's rule in the two-dimensional case can be expressed as
\begin{equation}
c = 2\left(2\pi a \xi_p \sqrt{\gamma^2 + \kappa^2} + 1\right)\xi_m.
\label{Carson}
\end{equation}

Avoiding overlap between adjacent harmonics requires $c < \xi_p$, which gives
\begin{equation*}
\xi_m < \frac{\xi_p}{4\pi a \xi_p \sqrt{\gamma^2 + \kappa^2} + 2} < \frac{\xi_p}{4\pi a \xi_p \sqrt{\gamma^2 + \kappa^2}}.
\end{equation*}
Consequently, an upper bound independent of $\xi_p$ is obtained: \begin{equation}
\label{eq:num_asymp}
\xi_m < \frac{1}{4\pi a \sqrt{\gamma^2 + \kappa^2}}.
\end{equation}
Normalizing $\xi_m$ by the upper bound gives the normalized frequency:
\[
\xi_n = 4\pi a \xi_m \sqrt{\gamma^2 + \kappa^2},
\]
which must lie between 0 and 1 to avoid spectral overlap.

Remarkably, for harmonics of order $(1,0)$ and $(0,1)$, this bound, derived from the condition of avoiding spectral overlap in the Fourier domain, coincides with the bound \eqref{eq:physical_const} obtained in the previous subsection, which arises from physical constraints, that is, the fact that the membrane cannot undergo folding or rupture.

This demonstrates that our Carson bound is tight for realistic deformations, i.e., those that do not exhibit folding or rupture.

In addition, to prevent aliasing and ensure correct phase unwrapping, the pattern frequency $\xi_p$ must satisfy Shannon’s sampling theorem, where $\xi_s$ is the spatial sampling frequency.

For a given harmonic $(\gamma,\kappa)$, using equation \eqref{eq:born_partial}, the maximum spatial frequency reached is
\[
\max(\gamma,\kappa)\, \xi_p \, \bigl(1 + 2\pi a \xi_m\bigr).
\]
Applying the bound from equation \eqref{eq:num_asymp} to simplify the expression by removing $\xi_m$ and $a$, we obtain an upper limit on the maximum frequency:
\[
\max(\gamma,\kappa)\, \xi_p \left(1 + \frac{1}{2\sqrt{\gamma^2+\kappa^2}}\right),
\]
which leads to the final constraint on the pattern frequency:
\begin{equation}
\label{eq:nup_asymp}
\xi_p < \frac{\sqrt{\gamma^2+\kappa^2}}{\max(\gamma,\kappa)(2\sqrt{\gamma^2+\kappa^2}+1)} \, \xi_s.
\end{equation}

This condition ensures that the wrapped phase variation between adjacent pixels remains below $\pi$, allowing unambiguous phase unwrapping and preventing from aliasing in the sampled 2D pattern.

From Eqs.~\eqref{eq:num_asymp} and \eqref{eq:nup_asymp}, the $(1,0)$ and $(0,1)$ harmonics are the most suitable choices, maximizing the usable bandwidth $\xi_m$ while allowing the highest aliasing-free frequency pattern $\xi_p$.

\subsection{Model for realistic observations}
In a realistic scenario, available datasets are digital images of the pattern \( I \) obtained through video acquisition and distortions introduced by the camera must be taken into account. Also potential imperfections of optics as well as measurement noise must be considered, leading to the following observation model for the modulated pattern $\tilde{I}_\textrm{d}$:
\begin{equation}
    \tilde{I}_\textrm{d} = I_\textrm{d} * \Psi + E,
    \label{eq:observation}
\end{equation}
where \( \Psi \) is the camera's point spread function (PSF), assumed to be spatially invariant and Gaussian-shaped, and \( E \) is centered white Gaussian noise with fixed variance \( \sigma^2 \). As expressed in \cite[Sec.~2.2]{delbracio2013two}, the noise in digital camera sensors is a mixture of luminance dependent (Poisson or photon counting) noise and luminance independent (Gaussian, thermal) noise, and is often approximated as white Gaussian noise. This justifies the noise model in our framework. The model based on expressions in \eqref{eq:num_asymp}, \eqref{eq:nup_asymp}, and \eqref{eq:observation}, is a realistic model for modulated patterns on membranes. The next section introduces a demodulation method allowing the local estimation of the parameters of the wave propagating through the membrane.

\section{Demodulation using the 2D analytic signal}
\label{sec:demod}
\subsection{Principle of the proposed estimator}

In this section, we introduce ADOPT (Analytical Demodulation of Periodic Texture), using the modulation model previously introduced, one can design a demodulation method based on the 2D analytic signal to estimate the in-plane waveform parameters. Specifically, we aim to provide an estimate \( (\hat{U}, \hat{V}) \) for the displacements \( (U, V) \) from the original pattern \( \tilde{I} \) and its modulated version \( \tilde{I}_\textrm{d} \). Our approach processes $\hat{U}$ and $\hat{V}$ separately. We detail below the estimation of $\hat{U}$ while $\hat{V}$ follows by the same lines.

 \begin{figure*}[ht]
     \centering
     \includegraphics[width=0.75\textwidth, clip]{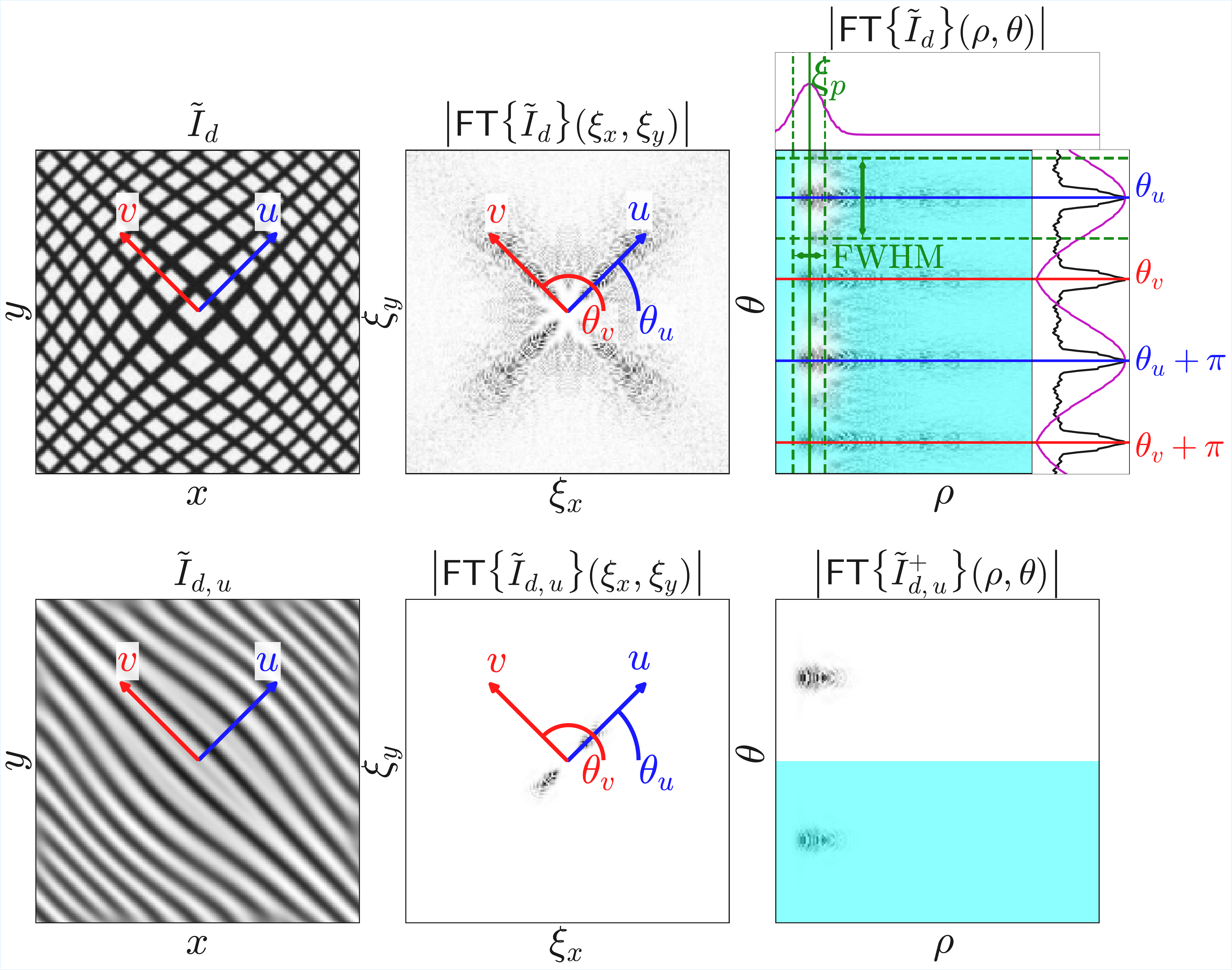}
     \caption{Orientation detection and filtering principle, where the regions highlighted in cyan are removed by filtering. \textbf{Top:} modulated pattern (left), its 2D Fourier transform in Cartesian (center) and polar coordinates with filters (right). \textbf{Bottom:} filtered image in the spatial domain (left), its Fourier transform (center) and the analytic signal spectrum in polar coordinates (right).}
     \label{fig:filter}
 \end{figure*}

 \begin{figure}[h!]
     \centering
     \includegraphics[width=9cm,clip]{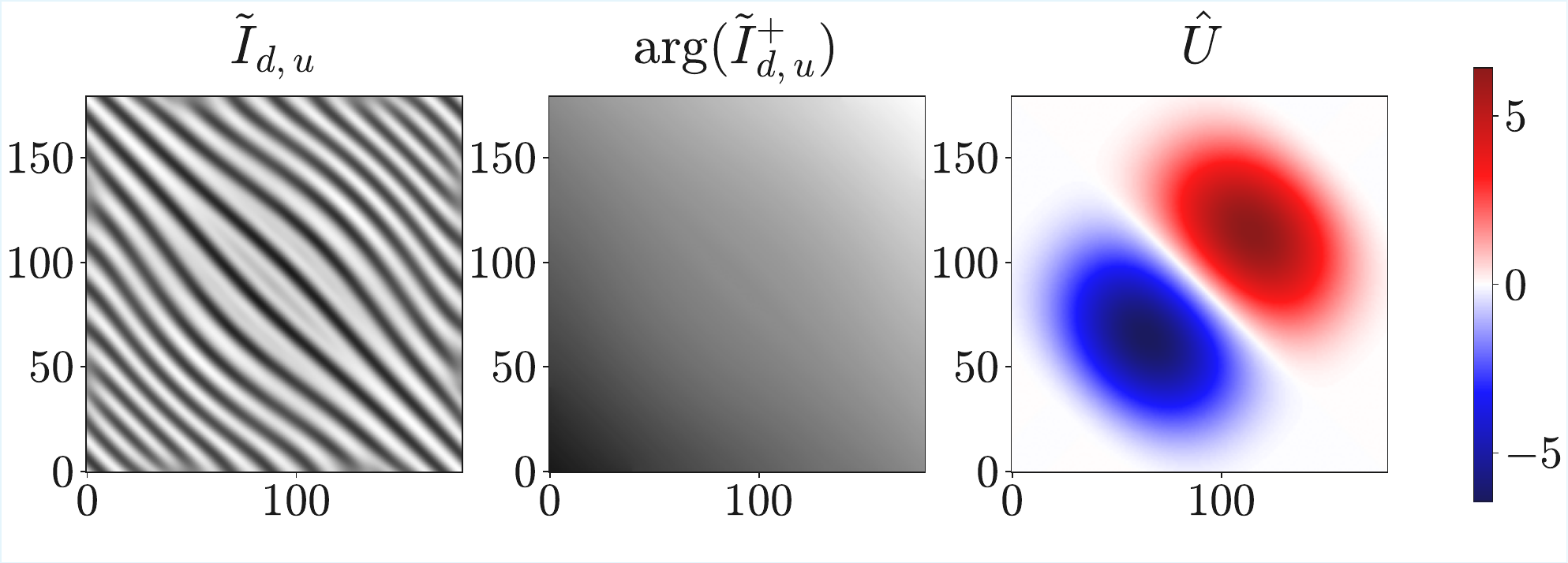}
     \caption{Displacement estimation of the $U$ component (in pixels). \textbf{Left:} filtered modulated pattern along $\theta_u$. \textbf{Center:} phase of the oriented analytic signal. \textbf{Right:} estimated displacement $\hat{U}$.}
     \label{fig:phase}
 \end{figure}

The pattern $\tilde{I}$ introduced previously exhibits a distinctive 2D Fourier transform (see Fig.~\ref{fig:pattern_FT}). The displacement estimation relies on computing an oriented 2D analytic signal on a filtered version of $\tilde{I
}_\textrm{d}$. The orientation is estimated first, then the phase is extracted. In general, the pattern contains two orientations $\theta_u$ and $\theta_v$. Those may not necessarily be perpendicular, though such a case minimizes spectral overlap. The angle $\theta_u$ defines the $u$-direction, along which lines orthogonal to this direction are regularly spaced, enabling the estimation of $U$. Likewise, $\theta_v$ defines the $v$-direction for estimating $V$.

The spectrum of $\tilde{I}_{\text{d}}$ (Fig.~\ref{fig:filter}, top middle) reveals two diagonal branches, associated to $\theta_u$ and $\theta_v$. In polar coordinates, orientations in $[0, 2\pi)$ appear as horizontal lines (Fig.~\ref{fig:filter}, top right), where the mean intensity along $\rho$, shown in black, exhibits peaks at $\theta_u$, $\theta_v$, $\theta_u + \pi$, and $\theta_v + \pi$.

In order to isolate $\theta_u$, Gaussian filters, shown in magenta, are applied along $\theta$ centered at $\theta_u$ and $\theta_u + \pi$, with the full width at half maximum (FWHM) of $\pi/2$ (shown in green). An additional Gaussian filter along $\rho$ retains only the first harmonic, centered at $\xi_p$ with FWHM of $\xi_p$. As a result, $\tilde{I}_{d,u}$ (Fig.~\ref{fig:filter}, bottom left and center) contains only the orientation $\theta_u$. Phase extraction is then performed with the partial oriented analytic signal \cite[Section 3.1.2]{bulow1999hypercomplex} aligned with $\theta_u$:
$$
   \mathsf{FT} \{ \tilde{I}^{+}_{d,u} \} = \mathsf{FT} \{ \tilde{I}_{d,u} \} \left( 1 + \text{sign}(\xi_x \cos \theta_u + \xi_y \sin \theta_u) \right),
$$
where $\text{sign}(x)$ returns $-1$, $0$, or $1$. In polar coordinates (Fig.~\ref{fig:filter}, bottom right), half of the spectrum $[\theta_u - \pi/2, \theta_u + \pi/2]$ is retained, and the rest set to zero. The inverse Fourier transform yields the complex analytic signal $\tilde{I}^{+}_{d,u}$, whose phase gives an estimate of the wrapped displacement with expression:
\begin{equation}
    \hat{U} = \frac{1}{2 \pi \xi_p}\arg \left<\tilde{I}^{+}_{ u},\tilde{I}^{+}_{d,u}\right>
\end{equation}
where $\langle a,b \rangle=\bar{a}b$ denotes the Hermitian inner product, and $\tilde{I}^{+}_{u}$ is obtained from the reference (non-deformed) image using the same procedure. The result is shown in Fig.~\ref{fig:phase} (right). To obtain a continuous displacement field, $\hat{U}$ is unwrapped using the phase unwrapping algorithm introduced in \cite{herraez2002fast}, which avoids the error propagation typical of line-by-line or column-by-column approaches. The same procedure is applied to compute $\hat{V}$, yielding the full 2D displacement field estimate $(\hat{U}, \hat{V})$.
This method can be regarded as an alternative formulation of the monogenic signal proposed by Felsberg \cite{felsberg2002low} and later extended by Di Zang \cite{zang2007signal}, where orientation is estimated using spherical harmonic filters before extracting the phase.

\subsection{Limit of resolution of the ADOPT estimator}

Having introduced the ADOPT (Analytical Demodulation Of Periodic Texture) estimator, we now investigate its intrinsic resolution limits. We consider an ideal scenario characterized by a high signal-to-noise ratio (SNR) and small displacements $(\delta U,\delta V)$.

Under these assumptions, the deformed image can be expressed as
\[
\tilde{I}_d = \tilde{I}(x - \delta U, y - \delta V).
\]
A first-order Taylor expansion, analogous to the classical optical flow formulation, yields
\[
\tilde{I}_d \approx \tilde{I}(x,y) - \tilde{I}_x \, \delta U - \tilde{I}_y \, \delta V,
\]
where $\tilde{I}_x = \frac{\partial \tilde{I}}{\partial x}$ and $\tilde{I}_y = \frac{\partial \tilde{I}}{\partial y}$ denote the spatial derivatives of the image. This leads to the optical flow constraint equation
\[
\tilde{I}_x \, \delta U + \tilde{I}_y \, \delta V = -\tilde{I}_t,
\]
where $\tilde{I}_t = \tilde{I}_d - \tilde{I}(x,y)$ represents the intensity change induced by the deformation. We denote $\nabla \tilde{I} = (\tilde{I}_x,\tilde{I}_y)$ and the Euclidean norm as $\|(a,b)\| = \sqrt{a^2 + b^2}$. The optical flow equation can thus be written as
\[
\langle (\delta U,\delta V), \nabla \tilde{I} \rangle = -\tilde{I}_t.
\]
Applying the Cauchy--Schwarz inequality yields
\[
\|(\delta U,\delta V)\| \, \|\nabla \tilde{I}\| \geq |\langle (\delta U,\delta V), \nabla \tilde{I} \rangle| = |\tilde{I}_t|.
\]
Therefore,
\[
\|(\delta U,\delta V)\| \geq \frac{|\tilde{I}_t|}{\|\nabla \tilde{I}\|}.
\]

The minimum detectable intensity variation is given by the quantization step, denoted by $\delta \tilde{I}$, thus $\min(|\tilde{I}_t|) = \delta \tilde{I}$. Therefore, in the limiting case, the minimum detectable displacement is then given by
\begin{equation}
    \|(\delta U,\delta V)\|_{\min} = \frac{\delta \tilde{I}}{\|\nabla \tilde{I}\|_{\max}}.
    \label{displ_min}
\end{equation}

For periodic pattern images $\tilde{I}$, the maximum spatial gradient magnitude is directly related to the spatial frequency $\xi_p$. Indeed, the periodic structure induces spectral components at harmonics of $\xi_p$, so the highest spatial frequency in the spectrum of $\tilde{I}$ is a multiple of $\xi_p$. The order of this highest harmonic can be inferred from the bandwidth of the PSF \( \Psi \) and the sampling frequency $\xi_s$. Thus, by Bernstein’s inequality, the spatial derivatives $\tilde{I}_x$ and $\tilde{I}_y$ are bounded by a term proportional to $\xi_p$, which leads to $\|\nabla \tilde{I}\|_{\max} \propto \xi_p$.
Consequently, the resolution limit satisfies
\[
\|(\delta U,\delta V)\|_{\min} \propto \frac{\delta \tilde{I}}{\xi_p}.
\]

This result highlights a fundamental property of the estimator: increasing the spatial frequency of the pattern improves the displacement resolution, provided that high-SNR conditions are maintained.
\section{Numerical Results and Computational Complexity}
\label{sec:results}
\subsection{Results and Comparison with the Cramér--Rao bound}

\begin{figure}[h!]
    \centering
    \includegraphics[width= 8.7cm]{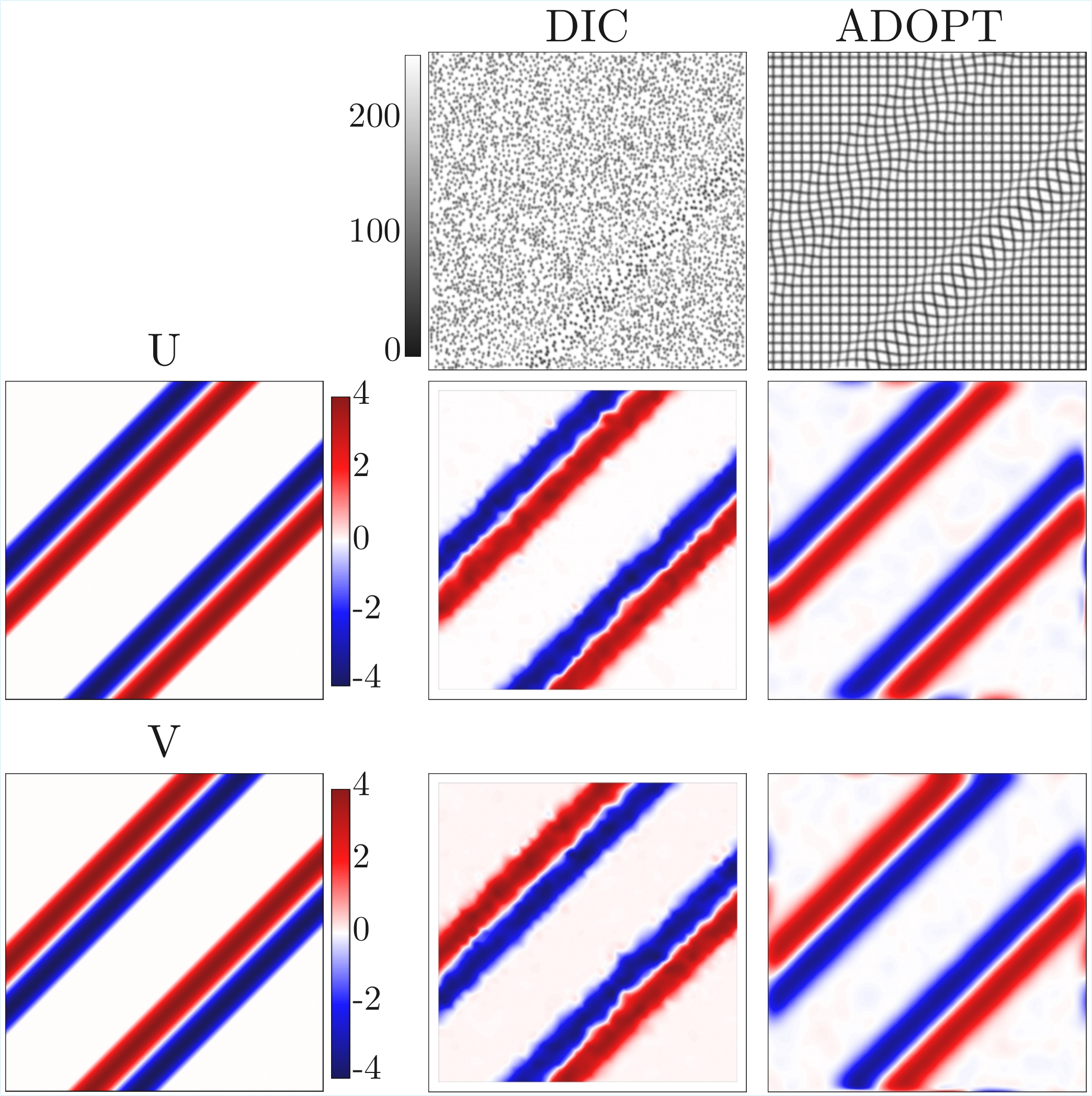}
    \caption{Wavefront reconstruction (\(U\) and \(V\) in pixels for SNR$=20$ dB). \textbf{Left column:} ground truth, \textbf{Middle column:} DIC (MSE = -11 dB), \textbf{Right column:} proposed method (MSE = -21 dB), DIC images are smaller due to edge interpolation limits.}
    \label{fig:results_dicvsours}
\end{figure}

We compare the proposed demodulation approach with the classical DIC technique \cite{rastogi2003photomechanics,grediac2012full}, which tracks waves in images using speckle patterns (Fig.~\ref{fig:results_dicvsours}, top middle). An example of wavefront estimation is given in Fig.~\ref{fig:results_dicvsours}. DIC interpolation and the limitations of its model introduce distortions and edge artifacts, whereas our method preserves the linear wavefront across the field.

For fairness, DIC hyperparameters are empirically optimized: the point pattern coverage is set to 30\%, and the correlation window size is adapted to the maximum displacement $a$. This main parameter is expressed as a percentage of the image size $\ell$, which constrains the wave's spatial frequency $\xi_m$ according to \eqref{eq:num_asymp}.

We choose the potentials $P_\alpha$, where $\alpha \in \{L,T\}$ (with $L$ for longitudinal and $T$ for transverse), and define them as functions of $(a,\xi_m,\theta_0,\phi_\alpha)$ and, implicitly, of $(x,y)$ through
$r_{\theta_0} = x \cos\theta_0 + y \sin\theta_0$:

\begin{equation}
P_\alpha(a,\xi_m,\theta_0,\phi_\alpha) = 
\frac{-a}{2\pi \xi_m} \, w(r_{\theta_0}, \phi_\alpha, \xi_m)\,
\cos\!\big( 2\pi \xi_m (r_{\theta_0} - \phi_\alpha) \big),
\end{equation}
with $w$ a Hanning window of length $1/\xi_m$, centered using the phase shift $\phi_\alpha$ and oriented along $\theta_0$, ensuring smooth apodization and alignment of the sinusoidal crest with the window center.
The resulting displacements are then given by \eqref{eq:UV}.

In our simulations, we consider a square image of size $400 \times 400$ pixels ($\ell = 400$ pixels). The maximum displacement is set to $a = 0.01\,\ell$ and $a = 0.005\,\ell$, the spatial frequency to $\xi_m = 3.33\,\ell^{-1}$, the propagation orientation to $\theta_0 = \pi/4$~rad, and the phase shifts to $\phi_L = 1$ and $\phi_T = 0.33$~rad. The signal-to-noise ratio (SNR) is fixed at 20~dB, and the pattern frequency is set to $\xi_p = 33.33\,\ell^{-1}$.
Realistic observations are then generated through the forward model described in Section~\ref{sec:mathmodel}.

Based on the Gaussian noise assumption, we numerically derive the Fisher information matrix over the parameters $(a,\xi_x,\theta_0,\phi_\alpha)$ using the Bhattacharyya–Slepian (Bangs–Slepian) formula \cite[Appendix 3C]{kay1993fund}, which in turn provides the Cramér–Rao bound (CRB) for the displacement estimator.

We also compute, via Monte-Carlo simulations, the mean squared error (MSE) of our displacement estimator. The parameter study is carried out by comparing the gap between the CRB and the empirical MSE curves for the different estimators.

\begin{figure}[h!]
    \centering
    \includegraphics[width= 8.7cm]{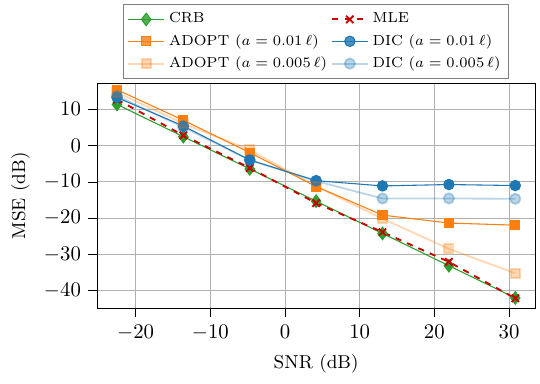}
    \caption{The averaged MSE of $\hat{U}$ and $\hat{V}$ versus SNR. Comparison between DIC, the proposed method, MLE, and the CRB.}
    \label{fig:resultat_DIC_Proposed}
\end{figure}

The behavior of the proposed estimator is illustrated in Fig.~\ref{fig:resultat_DIC_Proposed}. As the SNR increases, the MSE decreases and approaches the Cramér–Rao bound (CRB) before reaching a saturation plateau caused by spectral overlap. At low SNR, DIC can perform slightly better, as correlation-based estimators are less sensitive to additive noise. However, it saturates earlier due to its simplified local motion model, which does not capture intra-window nonlinear deformation. Above approximately 10~dB SNR, which corresponds to realistic experimental conditions, our method outperforms DIC.

The error behavior of DIC can be understood in two distinct regimes \cite[Sec.~6.3]{grediac2012full}. The first regime is associated with a noise-dominated behavior, in which the error decreases approximately linearly with increasing SNR. The second regime is related to the inability to fully describe the actual transformation within the correlation window. This error is governed by the deviation of the displacement field from a locally affine (or translational) behavior, and is therefore linked to both the maximum displacement amplitude and the spatial frequency content of the deformation.

In the first regime, the estimation error is dominated by additive noise. In this case, the mean-square error (MSE) scales proportionally to the variance of the noise divided by the squared magnitude of the image gradient, i.e.,
\[
\mathrm{MSE} \propto \frac{\sigma^2}{\|\nabla \tilde{I}\|^2},
\]
which is consistent with the classical formulation in \eqref{displ_min}. This behavior is expected since, at moderate to low SNR, the quantization level $\delta I$ becomes negligible compared to additive noise, which then constitutes the dominant source of uncertainty.

\begin{figure}[h!]
    \centering
    \includegraphics[width=9cm]{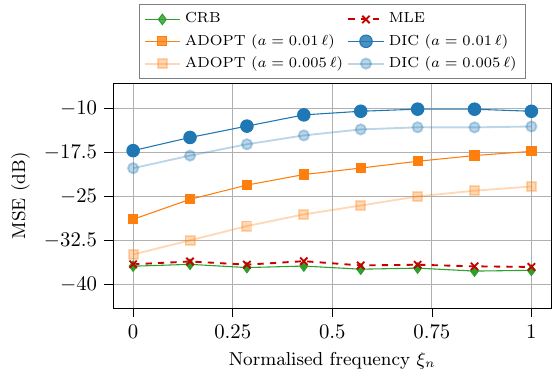}
    \caption{The averaged MSE of $\hat{U}$ and $\hat{V}$ versus normalized frequency. Comparison between DIC, the proposed method, MLE, and the CRB for SNR = 20 dB.}
    \label{fig:mse_vs_frequency}
\end{figure}

In a second regime, a noise-independent error floor appears, even when the SNR increases. This plateau is due to model mismatch: DIC assumes locally affine (purely translational) motion, whereas the true displacement field is not strictly translational within each subset. This induces a systematic bias that cannot be reduced by increasing SNR. This plateau decreases when the deformation amplitude $a$ decreases, as smaller deformations are closer to the local translation assumption. Conversely, increasing the spatial frequency enhances nonlinear effects within each window, which increases this bias, as illustrated in Fig.~\ref{fig:mse_vs_frequency}.

For our method ADOPT, a similar trend is observed. The estimation error decreases with increasing SNR until it reaches a plateau. This floor is mainly due to energy loss induced by filtering and finite-bandwidth reconstruction. Even when Carson’s rule is satisfied, a residual overlap between spectral components remains, leading to unavoidable information loss. This explains the observed saturation. This plateau decreases when the deformation amplitude $a$ decreases, since weaker modulations induce less spectral spreading (see \eqref{Carson}). Conversely, increasing the spatial frequency $\xi_p$ increases spectral overlap, which leads to higher error, as predicted by \eqref{Carson}.

Maximum likelihood estimation (MLE) provides a useful performance reference. Under the assumption of additive white Gaussian noise, the MLE consists in fitting a forward model of the deformed image to the observations by minimizing the squared difference between the measured image and its model-based reconstruction. In this framework, all parameters $(a,\xi_x,\theta_0,\phi_\alpha)$ are jointly estimated by solving a nonlinear optimization problem.

In the general case considered here, the observed pattern is not described by a simple parametric expression, which prevents deriving closed-form expressions for the derivatives of the log-likelihood. Nevertheless, a numerical approximation can be employed. In practice, the image model is interpolated using a differentiable scheme (e.g., bicubic interpolation), allowing the gradient of the log-likelihood with respect to the parameters to be approximated using finite differences. The Fisher information matrix can then be estimated numerically as the covariance of the score function, evaluated over multiple noise realizations (Monte Carlo simulations).

When the assumed model matches the true signal, the MLE is asymptotically efficient and achieves the Cramér--Rao bound (CRB), which explains its superior performance in Fig.~\ref{fig:resultat_DIC_Proposed}. However, this approach requires an accurate parametric description of the displacement field and relies on iterative optimization procedures, leading to a significantly higher computational cost and reduced robustness to model mismatch. As a result, this MLE-based approach is not readily generalizable to arbitrary waveforms and remains of limited relevance for real-world signals, unlike the proposed ADOPT method.

In addition, increasing the pattern frequency $\xi_p$ reduces the Cramér–Rao bound (CRB), thereby improving the theoretical estimation accuracy. For an unbiased estimator affected by additive white Gaussian noise with variance $\sigma^2$, the CRB is inversely proportional to the Fisher information, which measures the sensitivity of the observations to the parameter of interest. This Fisher information scales with the squared spatial gradient magnitude $\|\nabla \tilde{I}\|^2$. Consequently, the CRB decreases with increasing SNR, as the relative influence of noise decreases compared to signal variations.

Filtering improves robustness and helps maintain proximity to the CRB in the noise-dominated regime, but may increase the plateau at high SNR due to additional spectral attenuation.

DIC interpolation and the limitations of its model introduce distortions and edge artifacts, whereas our method preserves the linear wavefront across the field.

\subsection{Computational complexity}

Beyond estimation accuracy, computational efficiency is a critical factor for practical deployment. 
Classical DIC \cite{rastogi2003photomechanics,grediac2012full} relies on the evaluation of a similarity criterion over local blocks of size $b \times b$, combined with a search over a displacement domain of size $s \times s$. When the correlation is computed in the Fourier domain, each block correlation requires $\mathcal{O}(b^2 \log b^2)$ operations due to FFT-based convolution. This operation is repeated $s^2$ times over the search window of size $s \times s$, and applied over the entire image of size $\ell \times \ell$. The overall computational cost can therefore be approximated as
\[
\mathcal{O}\big(\ell^2 \, s^2 \, b^2 \log b^2\big).
\]

In contrast, the proposed method does not rely on local block matching or displacement search. The estimation is performed directly in the global Fourier domain, where the dominant cost is the computation of a small number of 2D FFTs over the entire image. This leads to a complexity of
\[
\mathcal{O}(\ell^2 \log \ell^2).
\]

We can see that increasing the block size or the interrogation window leads to a significant increase in computational complexity for DIC. In contrast, the proposed method remains computationally efficient, with a lower complexity, making it more suitable for practical deployment.

Overall, the proposed approach avoids both block-wise correlation and explicit search over the displacement domain, leading to a significantly reduced computational cost while preserving global consistency of the estimation.

\section{Experimental Validation}
\label{sec:exp}

In the following, we experimentally evaluate the suitability of the demodulation-based method described above (ADOPT) for characterizing the propagation of in-plane longitudinal waves in soft membranes under impulse excitation by comparing it to the standard 2D DIC analysis technique. To this end, the two methods are assessed and compared through a series of tests conducted on similar samples, sharing identical geometric and material properties, and subjected to the same vibration load.

\subsection{Experimental procedure} 

\begin{figure}[h!]
    \centering
    \includegraphics[width=4.5cm]{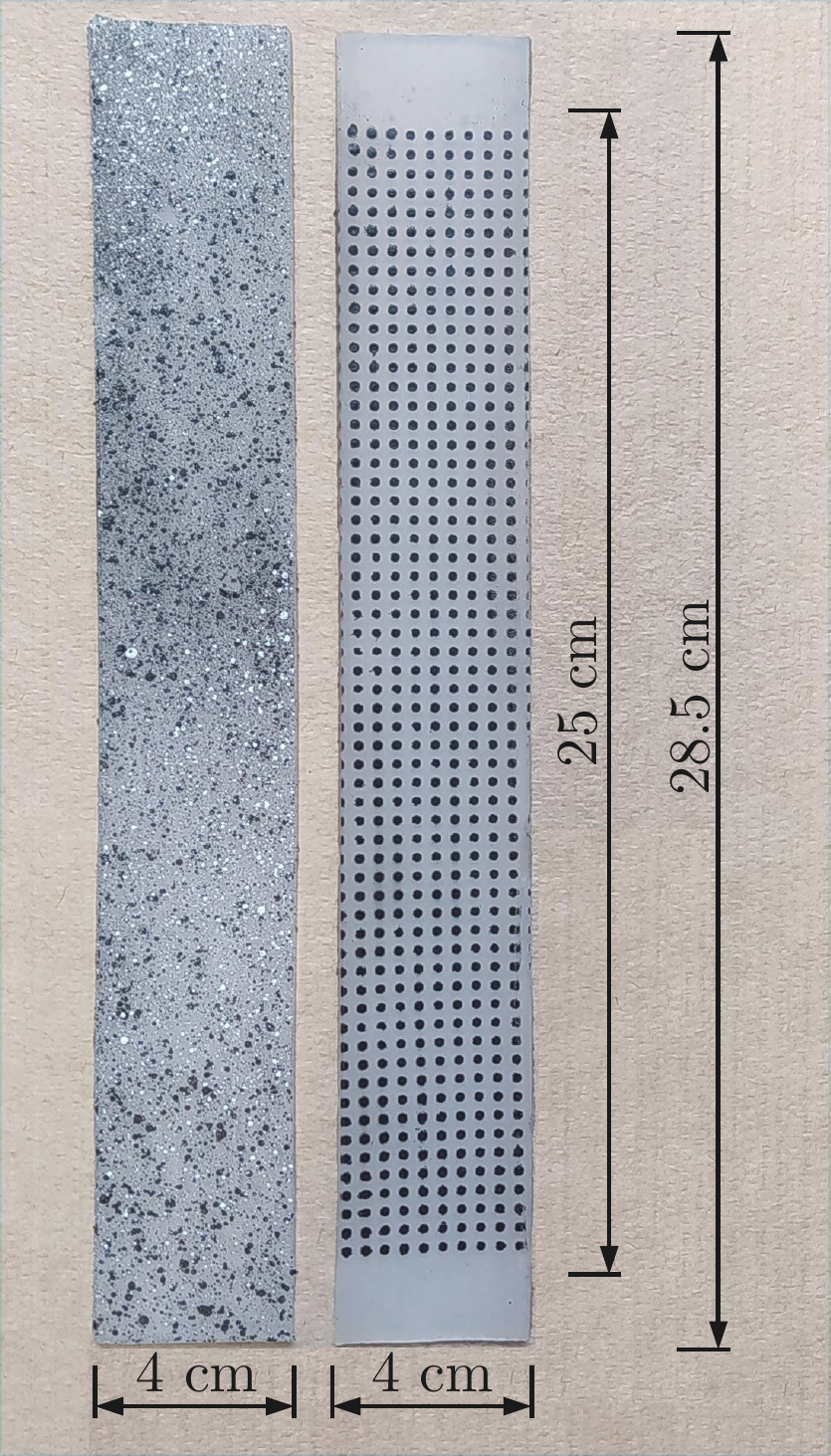}
    \caption{\textbf{Left:} silicone membrane with a speckle pattern. \textbf{Right:} silicone membrane with a periodic pattern.}
    \label{fig:echontillons}
\end{figure}

\emph{Materials --} A standard silicone elastomer (Smooth-on, Ecoflex\texttrademark~series, Shore hardness 00-30; \cite{lanoy2020dirac,luizard2023}) is selected to fabricate homogeneous membranes with controlled geometry and isotropic mechanical properties. Rectangular samples are processed with an initial length $\ell_0=$ 25~cm, a width $w_0=$ 4~cm, and a thickness $t_0=$ 3~mm at rest (Fig.~\ref{fig:echontillons}). All samples are fabricated using a dedicated Teflon\texttrademark~ mold, subjected to the same curing time (24~h), and maintained at the same ambient temperature (T~$\approx$ 25$^{\circ}$C) and relative humidity (RH~$\approx$ 45$\%$) until the impulse tests. Finally, surface patterns are spray-painted onto each strip to enable optical tracking of wave propagation throughout the test. Two types of patterns are employed, depending on the method selected for wave field measurements (Fig.~\ref{fig:echontillons})~: a random speckle pattern for DIC, and a periodic dot pattern for ADOPT. The latter design is produced using a customized stencil with evenly spaced circular holes (2~mm in diameter, with center-to-center spacing of 4~mm -- which defines the fundamental spatial period $d$ of the pattern). The choice of a periodic dot pattern rather than a grid, as used in numerical simulations for better visualization, is motivated by practical considerations. When the membrane is stretched, painted grid lines tend to crack or peel off, since the paint layer does not exhibit the same elasticity as the silicone substrate. In contrast, a dot pattern is more robust to deformations and better preserves the integrity of the surface texture.

\begin{figure}[h!]
    \centering
    \includegraphics[width=8cm]{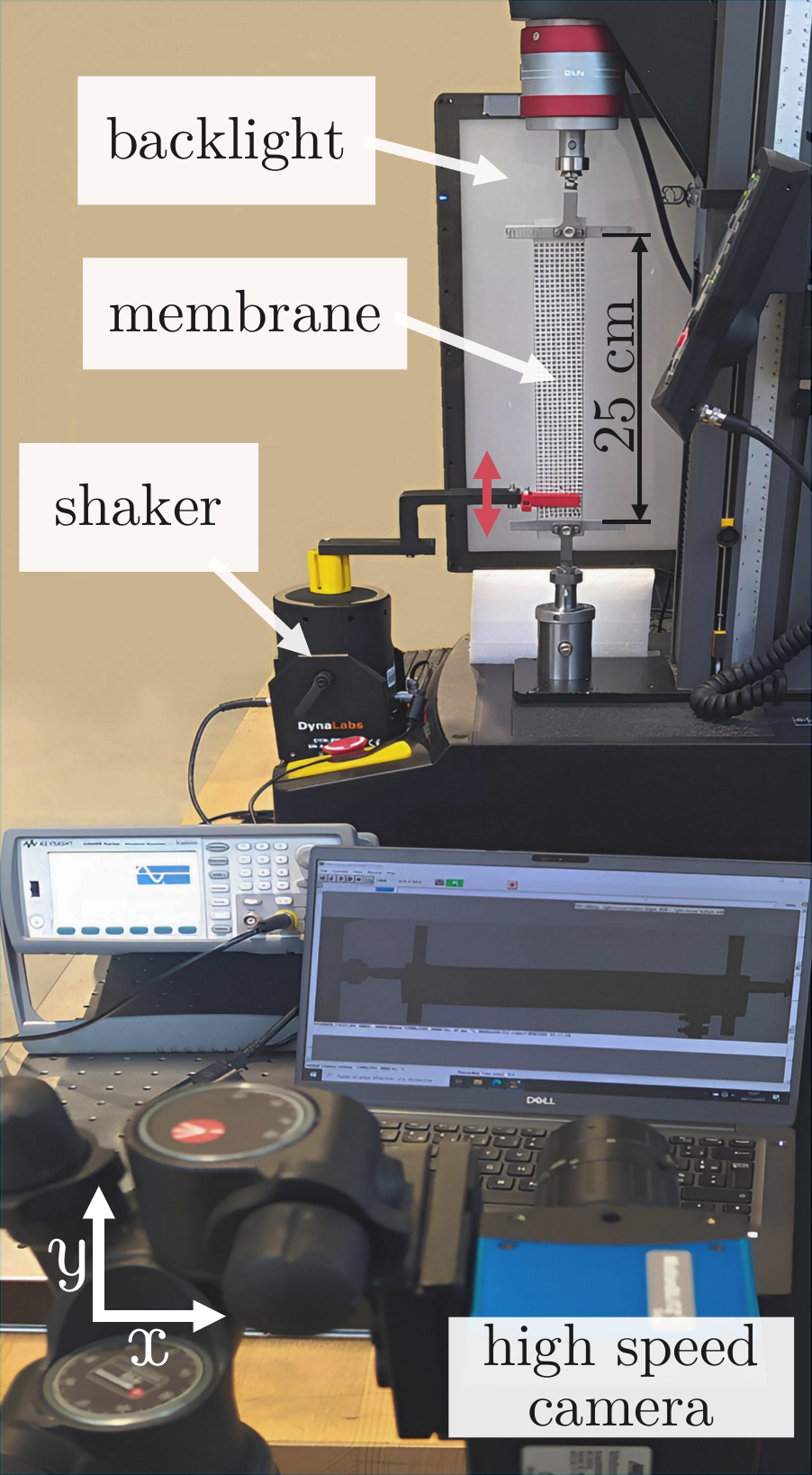}
    \caption{Experimental setup used to apply a vertical impulse (along the $y$-axis) excitation to the membrane using a shaker.}
    \label{fig:set_up}
\end{figure}

\emph{Overall setup and testing protocol --} As shown in Fig.~\ref{fig:set_up}, each membrane is clamped vertically using an electromechanical tension-compression testing machine (Instron$^{\mbox{\scriptsize{\textregistered}}}$~5944) equipped with a load cell of $\pm$ 100~N to ensure controlled placement, initial pre-load and boundary conditions. Customized textured grips were fabricated to facilitate the sample positioning and to minimize its slippage. An electrodynamic shaker (DynaLabs$^{\mbox{\scriptsize{\textregistered}}}$ DYN-PM-20) is mechanically coupled to the membrane via a rigid 3D-printed arm and clamp. It is driven by a Keysight 33500B series function generator, which produces an impulse excitation consisting of a single sinusoidal period set at 200~Hz. The shaker displacement can be adjusted using the shaker gain. Two configurations are considered: a low-gain configuration resulting in an excitation amplitude on the order of 22~$\mu$m, and a high-gain configuration resulting in an excitation amplitude on the order of 280~$\mu$m. A high-speed camera (Mikrotron$^{\mbox{\scriptsize{\textregistered}}}$ MotionBlitz$^{\mbox{\scriptsize{\textregistered}}}$ Cube4MGE-CM8) is used to visualize the resulting elastic waves traveling along the membrane. The membrane is back-illuminated using a high-intensity and uniform light source (Phlox LLUB LED-Backlight, 400 $\times$ 200 mm$^2$), exploiting its transparency to produce high-contrast images of the surface patterns. The images are acquired synchronously with the excitation and recorded at a rate of 3000 frames/s for a typical duration of 200 ms. The image size is 1280 $\times$ 386 pixels with a resolution of 30 pixels/cm. 
An experimental run involves selecting a membrane with a specific surface pattern and a specific impulse amplitude before proceeding with the excitation and image acquisition described above. Each run is repeated ten times.\\ 
This experimental setup allows a fair and realistic comparison between DIC and the proposed method under controlled conditions.\\

\emph{Data processing --} The recorded image sequences are then processed to quantify the spatio-temporal displacement fields induced by the waves in each case, and to derive the dispersion curves of the in-plane longitudinal waves. More specifically:
(i) membranes with periodic surface patterns are processed using the ADOPT algorithm;
(ii) DIC measurements are applied to membranes with random surface patterns, using a correlation window of size 32 $\times$ 32 pixels (one speckle is about 3 $\times$ 3 pixels), with an overlap of 90\% and bicubic interpolation.

\begin{figure}[h!]
    \centering
    \includegraphics[width=9cm]{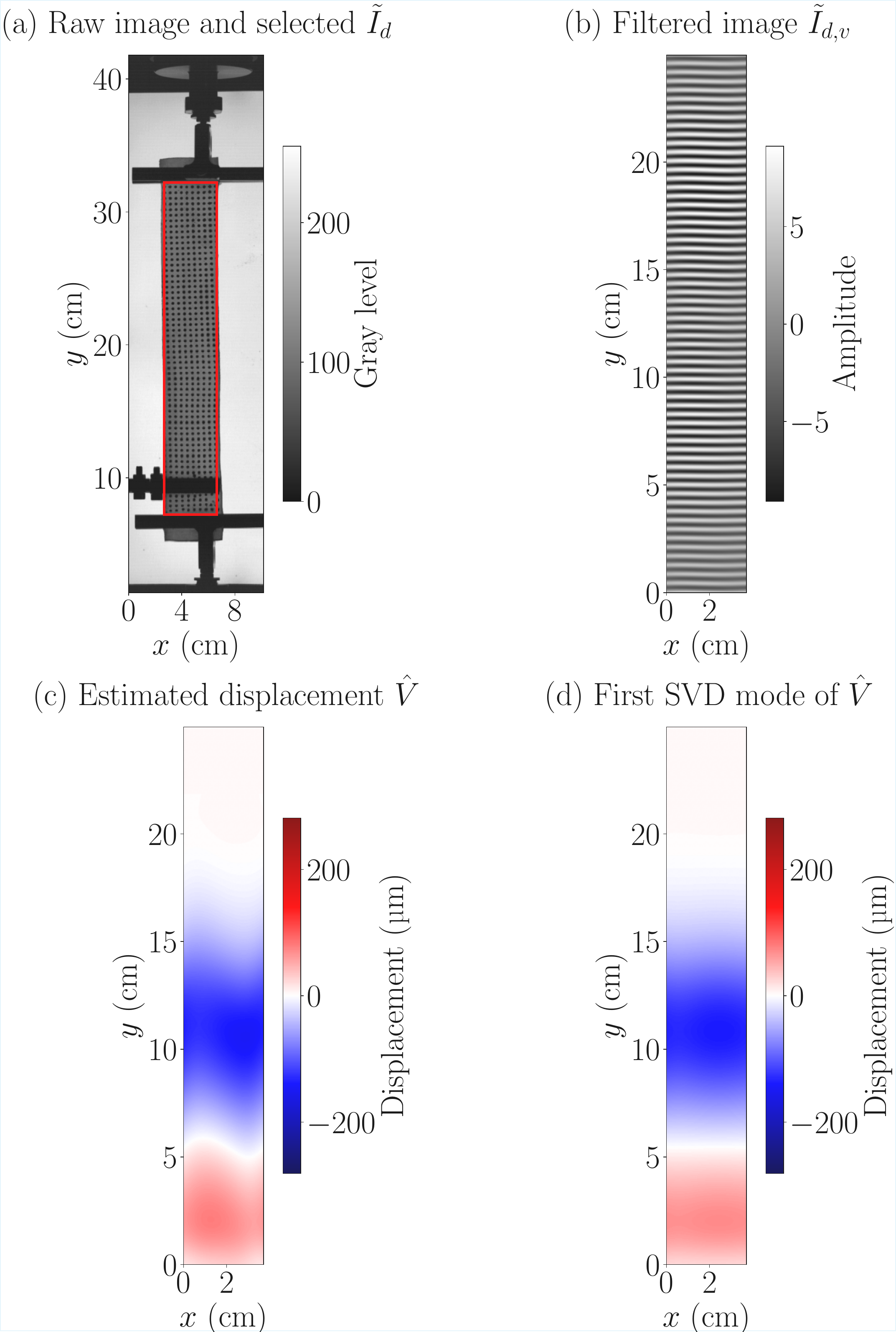}
    \caption{Estimation of the displacement along the $y$-axis from filtered patterns and SVD decomposition.}
    \label{fig:SVD}
\end{figure}

Starting from the image sequence acquired by the camera during excitation by the shaker, a spatio-temporal displacement map is constructed as follows. A reference unperturbed image $\tilde{I}$ is first selected, and all subsequent frames are processed relative to this reference. As illustrated in Fig.~\ref{fig:SVD}, the raw camera image is shown in the top-left panel, where the region of interest, i.e., the deformed membrane image $\tilde{I_d}$, is manually selected (red box). $\tilde{I_d}$ is spatially filtered along the $y$ direction (top right), yielding $\tilde{I_{d,v}}$, as described in the previous section. The proposed demodulation method is then applied to each filtered frame to estimate the in-plane displacement field, resulting in the displacement map shown in the bottom-left panel. A singular value decomposition (SVD) is subsequently performed on this field, and only the first mode is retained, providing a one-dimensional displacement profile (bottom right). This profile is further reduced to a 1D field by averaging along the $x$-axis (width), which suppresses transverse variations and emphasizes the displacement component along the propagation direction $y$. Repeating this procedure for all frames of the sequence and concatenating the resulting 1D profiles produces a spatio-temporal displacement representation, with spatial position (cm) on the vertical axis and time (ms) on the horizontal axis. The displacement colormap is expressed in $\mu\text{m}$.

As can be seen, we restrict ourselves to the displacement along the $y$-axis, meaning that we consider a 1D scenario. For the sake of simplicity, we deliberately neglect the displacement along the $x$-axis.

\subsection{Spatio-temporal maps and dispersion curve estimation}

\begin{figure}[h!]
    \centering
    \includegraphics[width=9cm]{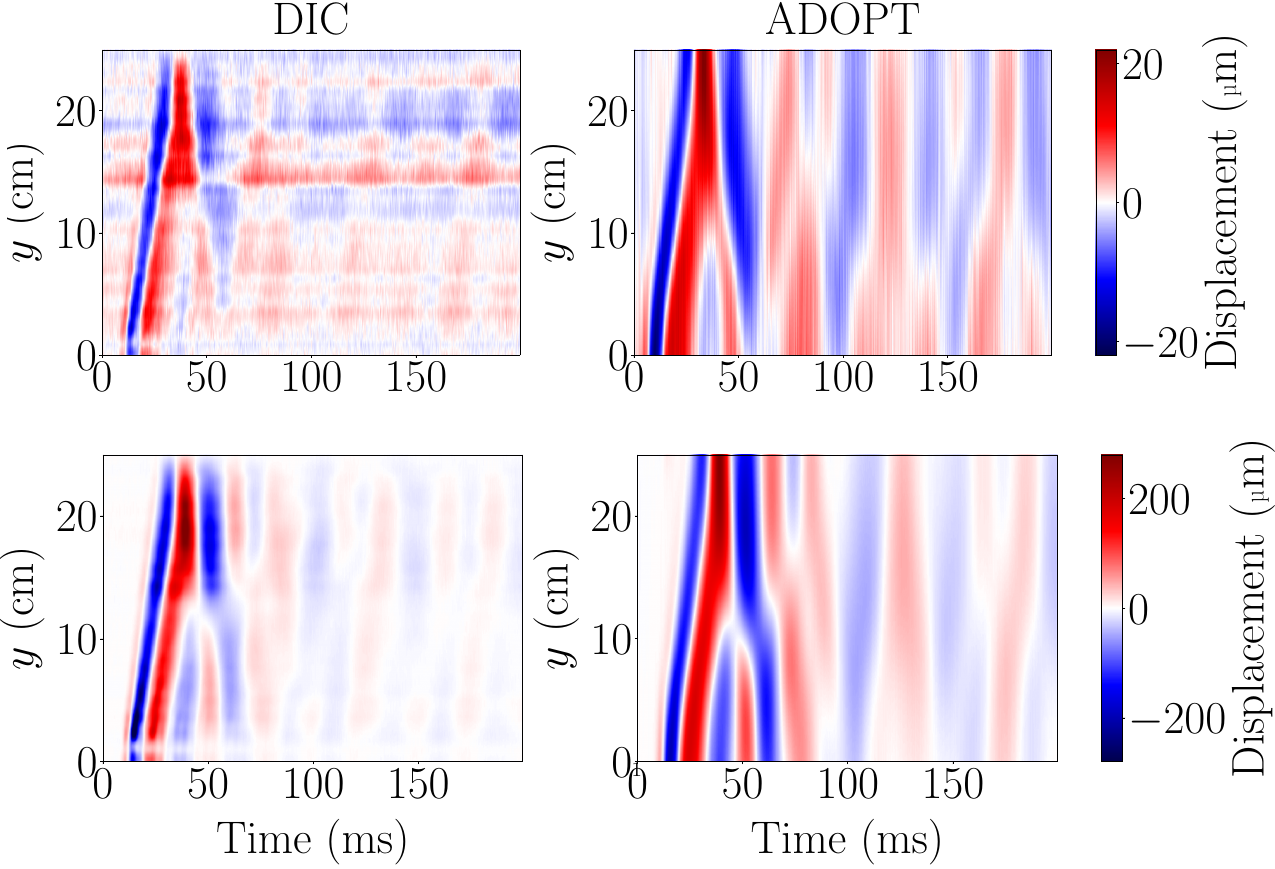}
    \caption{Resulting spatio-temporal displacement maps for DIC (left) and ADOPT (right) with \textbf{Top:} displacement amplitudes below 22 $\mu$m. \textbf{Bottom:} displacement amplitudes below 280 $\mu$m }
    \label{fig:exp_gain1_gain2}
\end{figure}

The resulting spatio-temporal maps, shown in Fig.~\ref{fig:exp_gain1_gain2}, reveal a wave propagating from the bottom of the image, where the excitation source (clamp connected to the shaker) is located. The wave travels upward, reflects at the upper boundary, and propagates back downward. This process repeats multiple times with progressively decreasing amplitude due to damping.

Comparing DIC and the proposed ADOPT method, as illustrated in the spatio-temporal figures, ADOPT consistently provides more accurate displacement reconstructions over the entire field of view. This improvement is particularly pronounced for small displacement amplitudes, where DIC becomes more sensitive to noise. In contrast, ADOPT maintains stable performance.

The objective is to exploit these displacement maps to estimate the dispersion curves of longitudinal waves using both methods (ADOPT and DIC) and for both excitation levels. Each experiment is repeated ten times in order to assess the variability of the estimated dispersion curves.

\begin{figure}[h!]
    \centering
    \includegraphics[width=9cm]{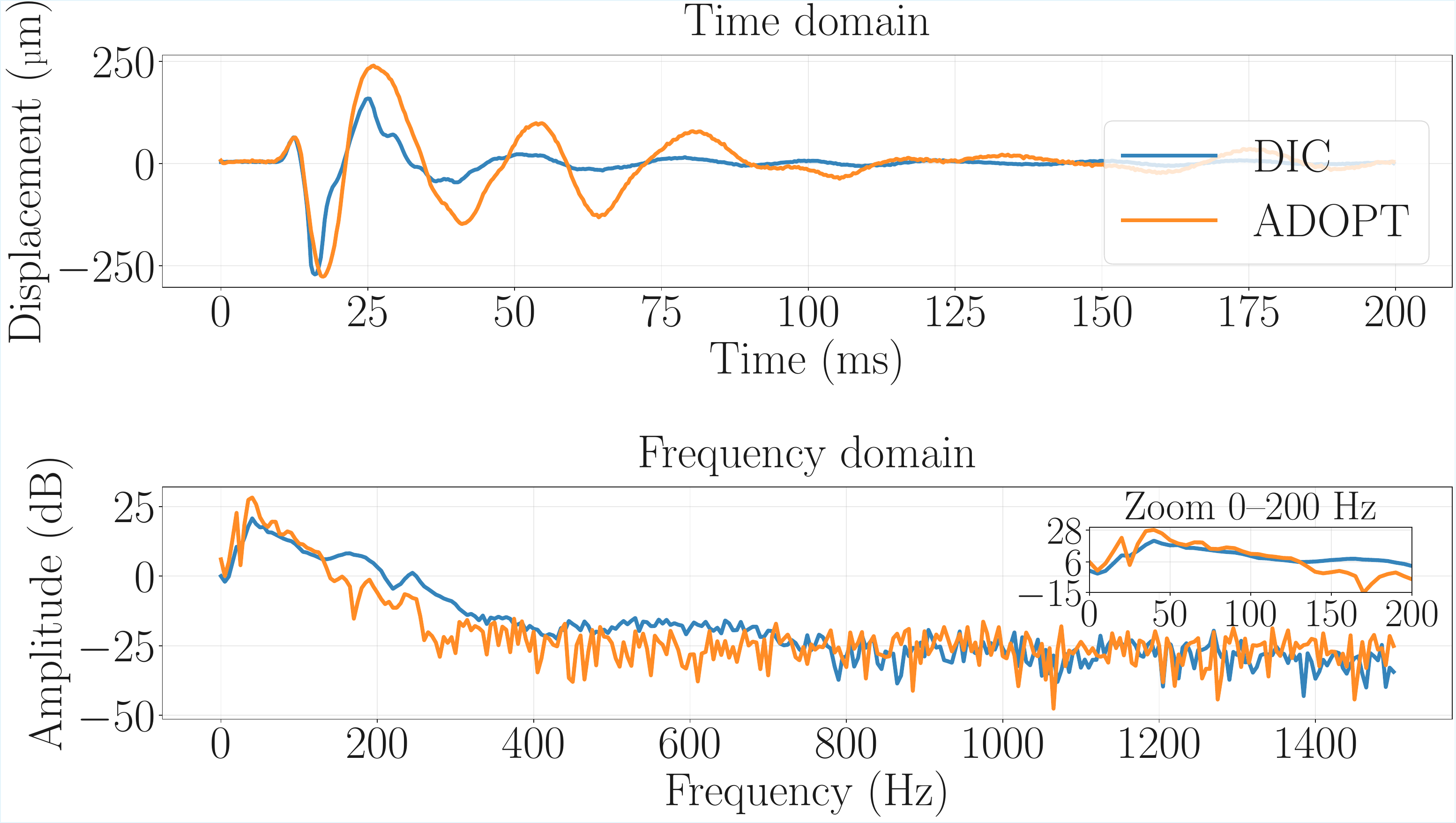}
    \caption{Time-domain and frequency-domain representations of the signal extracted along the $y=2$ cm line from the spatio-temporal map (amplitudes below 280 $\mu$m), corresponding to the clamp motion and thus the shaker motion, estimated using DIC and ADOPT.}
    \label{fig:DIC_pot}
\end{figure}

To determine the relevant frequency range for the analysis, the displacement at the excitation point is first examined. This is achieved by extracting the signal at a spatial position of $2~\text{cm}$, corresponding to the clamp location. Figure~\ref{fig:DIC_pot} shows the estimated motion of the shaker for displacement amplitudes below 280 $\mu$m, in both the time and frequency domains. The estimation obtained with the proposed method appears smoother. The corresponding frequency response indicates that the energy of the excitation rapidly decreases outside a limited frequency band. In particular, a drop of approximately $20~\text{dB}$ is observed outside the interval $40$--$80~\text{Hz}$. This limitation is inherent to the shaker, whose bandwidth constrains the frequency content of the generated impulse, as can also be verified from the manufacturer specifications.

Consequently, the frequency range of interest is restricted to $40$--$80~\text{Hz}$, where most of the injected mechanical energy is concentrated and where the signal-to-noise ratio remains sufficiently high for reliable estimation.

Dispersion curves are then estimated using a propagative approach. The incident wave traveling from the excitation source toward the opposite boundary is first isolated. For each frequency bin, the phase of the spatial Fourier transform is extracted as a function of position. This phase exhibits an approximately linear evolution along the propagation direction, allowing the wavenumber $k$ to be estimated via linear regression. The phase velocity is then obtained from the relation $c = 2\pi f / k$. Repeating this procedure over the frequency range of interest provides the dispersion curve.

\begin{figure}[h!]
    \centering
    \includegraphics[width=9cm]{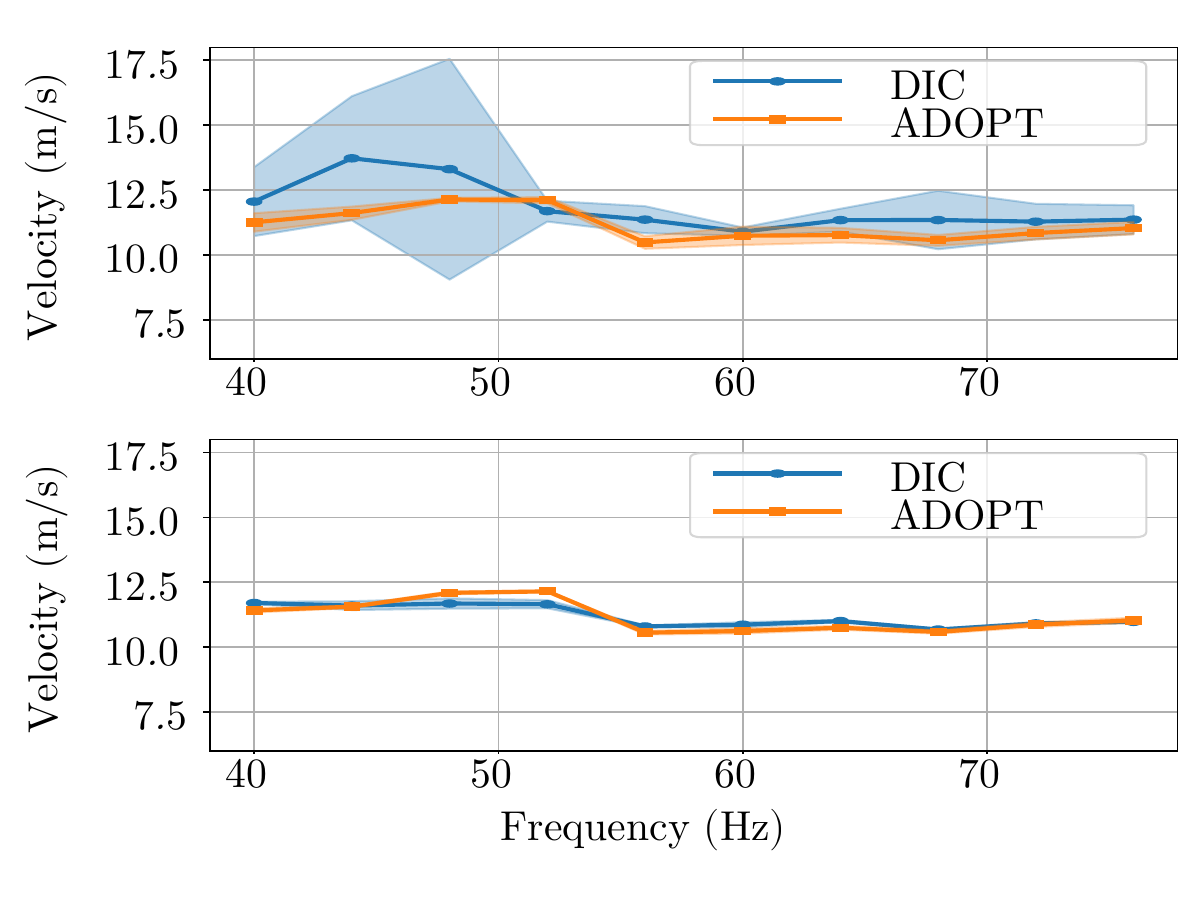}
    \caption{Estimated dispersion curves. \textbf{Top:} displacement amplitudes below 22 $\mu$m. \textbf{Bottom:} displacement amplitudes below 280 $\mu$m. The solid line shows the mean curve, and the shaded region corresponds to ± one standard deviation.}
    \label{fig:disp_gain_1_gain_2}
\end{figure}

The resulting dispersion curves are shown in Fig.~\ref{fig:disp_gain_1_gain_2}. These results highlight the superiority of the proposed method in experimental conditions. In particular, the variance of the DIC-based estimates is significantly larger in the low-displacement regime (below 22 $\mu$m), whereas the proposed method exhibits consistent performance across both excitation levels.
For low frequencies (between 40 and 55~Hz), corresponding to wavelengths $\lambda$ on the order of 27.5 to 20~cm (using $\lambda = c/f$ with $c \approx 11$~m/s), a slight increase in the wave speed may be observed in the dispersion curve. This behavior can be attributed to the finite size of the specimen, as the wavelength becomes comparable to its length $\ell_0=$ (25~cm), leading to interference between incident and reflected waves. As a result, the measured field departs from a purely propagative regime, introducing a bias in the phase-based velocity estimation. Beyond 55~Hz, the estimated dispersion stabilizes, consistent with the expected behavior of longitudinal waves in this frequency range, which are essentially non-dispersive.

The obtained wave speed remains physically consistent and agrees well with values reported in the literature for similar materials \cite{delory2024viscoelastic, zhu2025young}.

These experimental results confirm the observations made in simulation. While DIC exhibits large variance and inaccurate dispersion estimates for small displacements, converging only when deformation amplitudes become sufficiently large, the proposed method remains stable and consistent over the entire displacement range. This behavior is reflected in the dispersion curves, where DIC produces scattered estimates at low amplitudes, whereas the proposed approach yields smooth and repeatable results. These findings demonstrate that periodic patterns combined with the proposed demodulation framework provide superior performance compared to DIC for wave characterization on soft membranes.

\section{Conclusion}

This work addressed the problem of non-contact estimation of in-plane wave fields from image sequences by exploiting periodic surface patterns. A physics-informed modulation model was first introduced to describe the effect of wave-induced displacements on periodic textures, together with theoretical constraints ensuring physical consistency and preventing spectral overlap. Based on this model, a demodulation framework named ADOPT (Analytical Demodulation Of Periodic Texture) was developed, relying on an oriented two-dimensional analytic signal to jointly estimate displacement components.

The proposed approach was first assessed theoretically through the Cramér--Rao bound and benchmarked against Digital Image Correlation (DIC), which remains a standard reference method. Numerical simulations showed that ADOPT achieves near-optimal performance in the high-SNR regime and significantly outperforms DIC, particularly for small displacement amplitudes, where DIC exhibits increased variance and sensitivity to model mismatch. In addition, the proposed method reduces computational cost by avoiding local correlation and search procedures.

Experimental validation on silicone membranes confirmed these findings. Dispersion curves of longitudinal waves were successfully retrieved under different excitation levels, demonstrating that ADOPT provides stable and repeatable estimates over the full displacement range, whereas DIC only converges reliably for sufficiently large motions. The estimated wave speeds were found to be physically consistent and in good agreement with values reported in the literature for similar viscoelastic materials.

Overall, the results demonstrate that combining periodic surface patterns with a model-based demodulation strategy offers a robust alternative to classical DIC for wave characterization in soft media, which is particularly interesting for tracking small-amplitude deformations.

Future work will focus on extending the approach to out-of-plane displacement estimation. The method will also be applied to the characterization of highly deformable materials such as textile fabrics, which exhibit complex deformation fields.

\section*{Acknowledgments}
This work is funded by the 2024 UGA IRGA program through the IP-OP project. The 3SR Lab is part of the LabEx Tec 21 (Investissements d'Avenir - Grant Agreement No. ANR-11-LABX-0030) and the Carnot PolyNat Institute (Investissements d'Avenir - Grant Agreement No. ANR-
16-CARN-0025-01). We would like to thank François Bonnel (CMTC, Grenoble INP - UGA) for this helpful assistance with the optical setup for high-speed recordings, as well as Jean-Baptiste Chabert (Assistant Engineer, Univ. Grenoble Alpes, 3SR Lab)  for his help in designing experimental 3D-printed components.

\section*{Authors declarations}
The authors have no conflicts of interest to disclose.
\section*{Data availability}
The data generated and/or analyzed during the current study are available from the corresponding author upon request.


\bibliographystyle{unsrt}
\bibliography{sampbib}

@book{kay1993fund,
author = {Kay, Steven M.},
title = {Fundamentals of statistical signal processing: estimation theory},
year = {1993},
isbn = {0133457117},
publisher = {Prentice-Hall, Inc.},
address = {USA}
}

@book{graff2012wave,
  title={Wave motion in elastic solids},
  author={Graff, Karl F},
  year={2012},
  publisher={Courier Corporation}
}

@book{pinsky2008introduction,
  title={Introduction to Fourier analysis and wavelets},
  author={Pinsky, Mark A},
  volume={102},
  year={2008},
  publisher={American Mathematical Soc.}
}

@article{lanoy2020dirac,
  title={Dirac cones and chiral selection of elastic waves in a soft strip},
  author={Lanoy, Maxime and Lemoult, Fabrice and Eddi, Antonin and Prada, Claire},
  journal={Proceedings of the National Academy of Sciences},
  volume={117},
  number={48},
  pages={30186--30190},
  year={2020},
  publisher={National Acad Sciences}
}

@article{delory2024viscoelastic,
  title={Viscoelastic dynamics of a soft strip subject to a large deformation},
  author={Delory, Alexandre and Kiefer, Daniel A and Lanoy, Maxime and Eddi, Antonin and Prada, Claire and Lemoult, Fabrice},
  journal={Soft Matter},
  volume={20},
  number={9},
  pages={1983--1995},
  year={2024},
  publisher={Royal Society of Chemistry}
}

@phdthesis{felsberg2002low,
  title={Low-level image processing with the structure multivector},
  author={Felsberg, Michael},
  school={Kiel University},
  year={2002}
}

@phdthesis{bulow1999hypercomplex,
  title={Hypercomplex spectral signal representations for the processing and analysis of images},
  author={B{\"u}low, Thomas},
  school={Kiel University},
  year={1999}
}

@article{zang2007signal,
  title={Signal modeling for two-dimensional image structures},
  author={Zang, Di and Sommer, Gerald},
  journal={Journal of Visual Communication and Image Representation},
  volume={18},
  number={1},
  pages={81--99},
  year={2007},
  publisher={Elsevier}
}

@article{herraez2002fast,
  title={Fast two-dimensional phase-unwrapping algorithm based on sorting by reliability following a noncontinuous path},
  author={Herr{\'a}ez, Miguel Arevallilo and Burton, David R and Lalor, Michael J and Gdeisat, Munther A},
  journal={Applied optics},
  volume={41},
  number={35},
  pages={7437--7444},
  year={2002},
  publisher={Optica Publishing Group}
}

@article{wildeman2018real,
  title={Real-time quantitative Schlieren imaging by fast Fourier demodulation of a checkered backdrop},
  author={Wildeman, Sander},
  journal={Experiments in Fluids},
  volume={59},
  number={6},
  pages={97},
  year={2018},
  publisher={Springer}
}

@book{grediac2012full,
  title={Full-field measurements and identification in solid mechanics},
  author={Gr{\'e}diac, Michel and Hild, Fran{\c{c}}ois},
  year={2012},
  publisher={John Wiley \& Sons}
}

@book{rastogi2003photomechanics,
  title={Photomechanics},
  author={Rastogi, Pramod K},
  volume={77},
  year={2003},
  publisher={Springer Science \& Business Media}
}

@article{carson2006notes,
  title={Notes on the theory of modulation},
  author={Carson, John R},
  journal={Proceedings of the institute of radio engineers},
  volume={10},
  number={1},
  pages={57--64},
  year={2006},
  publisher={IEEE}
}

@book{brodatz1966textures,
  title     = {Textures: a photographic album for artists and designers},
  author    = {Brodatz, Phil},
  year      = {1966},
  address   = {New York},
  publisher = {Dover Publications, Inc.}
}

@article{luizard2023,
  title={Flow-induced oscillations of vocal-fold replicas with tuned extensibility and material properties},
  author={Luizard, Paul and Bailly, Lucie and Yousefi-Mashouf, Hamid and Girault, Rapha{\"e}l and Org{\'e}as, Laurent and Henrich Bernardoni, Nathalie},
  journal={Scientific reports},
  volume={13},
  number={1},
  pages={22658},
  year={2023},
  publisher={Nature Publishing Group UK London}
}

@article{chasiotis2002new,
  title={A new microtensile tester for the study of MEMS materials with the aid of atomic force microscopy},
  author={Chasiotis, Ioannis and Knauss, Wolfgang G},
  journal={Experimental Mechanics},
  volume={42},
  number={1},
  pages={51--57},
  year={2002},
  publisher={Springer}
}

@article{leprince2007automatic,
  title={Automatic and precise orthorectification, coregistration, and subpixel correlation of satellite images, application to ground deformation measurements},
  author={Leprince, Sbastien and Barbot, Sylvain and Ayoub, Franois and Avouac, Jean-Philippe},
  journal={IEEE Transactions on Geoscience and Remote Sensing},
  volume={45},
  number={6},
  pages={1529--1558},
  year={2007},
  publisher={IEEE}
}

@article{besnard2010characterization,
  title={Characterization of necking phenomena in high-speed experiments by using a single camera},
  author={Besnard, Gilles and Lagrange, Jean-Michel and Hild, Fran{\c{c}}ois and Roux, St{\'e}phane and Voltz, Christophe},
  journal={EURASIP Journal on Image and Video Processing},
  volume={2010},
  number={1},
  pages={215956},
  year={2010},
  publisher={Springer}
}

@article{sutton2008strain,
  title={Strain field measurements on mouse carotid arteries using microscopic three-dimensional digital image correlation},
  author={Sutton, MA and Ke, X and Lessner, SM and Goldbach, M and Yost, M and Zhao, F and Schreier, HW},
  journal={Journal of Biomedical Materials Research Part A: An Official Journal of The Society for Biomaterials, The Japanese Society for Biomaterials, and The Australian Society for Biomaterials and the Korean Society for Biomaterials},
  volume={84},
  number={1},
  pages={178--190},
  year={2008},
  publisher={Wiley Online Library}
}

@article{zhu2025young,
  title={Young's modulus estimation of a soft viscoelastic rod using optical elastography},
  author={Zhu, Jiayuan and Legrand, Fran{\c{c}}ois and Gr{\'e}goire, Sibylle and Torres, Jorge and Catheline, Stefan and Giammarinaro, Bruno},
  journal={The Journal of the Acoustical Society of America},
  volume={158},
  number={4},
  pages={3203--3209},
  year={2025},
  publisher={AIP Publishing}
}

@article{alfarano2024estimating,
  title={Estimating optical flow: A comprehensive review of the state of the art},
  author={Alfarano, Andrea and Maiano, Luca and Papa, Lorenzo and Amerini, Irene},
  journal={Computer vision and image understanding},
  volume={249},
  pages={104160},
  year={2024},
  publisher={Elsevier}
}

@phdthesis{delbracio2013two,
  title={Two Problems of Digital Image Formation: Recovering the Camera Point Spread Function and Boosting Stochastic Renderers by Auto-similarity Filtering},
  author={Delbracio, Mauricio},
  year={2013},
  school={{\'E}cole normale sup{\'e}rieure de Cachan-ENS Cachan}
}

@article{domino2016faraday,
  title={Faraday wave lattice as an elastic metamaterial},
  author={Domino, L and Tarpin, M and Patinet, Sylvain and Eddi, A},
  journal={Physical Review E},
  volume={93},
  number={5},
  pages={050202},
  year={2016},
  publisher={APS}
}

@article{grasland2018ultrafast,
  title={Ultrafast imaging of cell elasticity with optical microelastography},
  author={Grasland-Mongrain, Pol and Zorgani, Ali and Nakagawa, Shoma and Bernard, Simon and Paim, Lia Gomes and Fitzharris, Greg and Catheline, Stefan and Cloutier, Guy},
  journal={Proceedings of the National Academy of Sciences},
  volume={115},
  number={5},
  pages={861--866},
  year={2018},
  publisher={National Academy of Sciences}
}

\end{document}